\documentclass[10pt,aps,prd,showpacs,superscriptaddress,twocolumn]{revtex4}

\usepackage{amsfonts}
\usepackage{amssymb}
\usepackage{amsmath}
\usepackage{MnSymbol}
\usepackage{graphicx}

\newcommand{\ra}{\mathsf{R}}
\newcommand{\su}{\mathcal{S}}

\date{\today}

\begin{document}

\title{Instability of nonminimally coupled scalar fields in 
the spacetime of slowly rotating compact objects}

\author{Raissa F. P. Mendes}
\email{rfpm@ift.unesp.br}
\affiliation{Instituto de F\'\i sica Te\'orica, Universidade Estadual Paulista,
Rua Dr. Bento Teobaldo Ferraz 271, 01140-070, S\~ao Paulo, S\~ao Paulo, Brazil}

\author{George E. A. Matsas}
\email{matsas@ift.unesp.br}
\affiliation{Instituto de F\'\i sica Te\'orica, Universidade Estadual Paulista,
Rua Dr. Bento Teobaldo Ferraz 271, 01140-070, S\~ao Paulo, S\~ao Paulo, Brazil}

\author{Daniel A. T. Vanzella}
\email{vanzella@ifsc.usp.br}
\affiliation{Instituto de F\'\i sica de S\~ao Carlos,
Universidade de S\~ao Paulo, Caixa Postal 369, 13560-970, 
S\~ao Carlos, S\~ao Paulo, Brazil}

\begin{abstract}
Nonminimally coupled free scalar fields may be unstable in the
spacetime of compact objects. Such instability can be triggered 
by classical seeds or, more simply, by quantum 
fluctuations giving rise to the so-called {\em vacuum awakening 
effect}. Here, we investigate how the parameter space which 
characterizes the instability is affected when the object gains 
some rotation. For this purpose, we focus on the stability 
analysis of nonminimally coupled scalar fields in the spacetime 
of slowly spinning matter shells.
\end{abstract}

\pacs{04.62.+v, 04.40.Dg, 95.30.Sf}

\maketitle

\section{Introduction}
\label{sec:introduction}

In Refs.~\cite{lv10,lmv10} it was found that quantum 
fluctuations of certain nonminimally coupled free scalar 
fields defined in the spacetime of some relativistic stars 
can undergo an exponential amplification in time (see 
also Ref.~\cite{llmv12} for a comprehensive discussion). 
This ``vacuum awakening effect'' can be seen as the quantum
counterpart of the classical linear instability experienced
by these nonminimally coupled fields in such spacetimes 
\cite{mmv13}, or, more generally, of the classical 
instability observed in certain scalar-tensor theories 
\cite{Harada97-98,Cardoso2013}. 

A particularly interesting implication of this instability 
(neglecting restabilization mechanisms 
\cite{Pani2011}) is the possibility of ruling out certain 
classes of nonminimally coupled scalar fields by, e.g.,
determining the mass-to-radius ratio of relativistic stars 
with known equations of state. For this purpose, it is 
interesting to allow for natural deviations of the symmetry
assumptions imposed on the stellar models considered in 
Ref.~\cite{lmv10}, such as spherical symmetry and staticity, 
and investigate whether the conclusions would change 
significantly. This was partially done in Ref.~\cite{lmmv13}, 
where a class of static spheroidal shells was taken as the 
source of the gravitational field, and it was shown how the 
space of parameters that trigger the instability changes when
increasingly higher deviations from spherical symmetry are
considered. The aim of the present paper is to complement that
analysis by studying the effects of rotation, which is an
ubiquitous and often important property of astrophysical compact
objects such as neutron stars, whose spin frequency can be as 
high as 700 Hz~\cite{Pulsar}.

We begin, in Sec.~\ref{sec:quantization}, by discussing some 
aspects of the quantization of nonminimally coupled free scalar
fields containing unstable modes in a background which is flat 
in the asymptotic past and stationary and axially symmetric
in the future. In Sec.~\ref{sec:1storder} we present a simple
general argument that shows that the parameter space which
characterizes the instability is not modified at first order 
in the compact object's angular momentum. Then, we investigate
second order deviations from staticity in a particular model,
taking as the source of the gravitational field a class of 
slowly spinning shells. The general properties of the shell 
spacetime are presented in Sec.~\ref{sec:spacetime}.
Considering spinning thin shells allows us to push the analytical
treatment further and arrive at clear conclusions about the 
role played by rotation on the instability. This is pursued in 
Sec.~\ref{sec:2ndorder}. Section~\ref{sec:conclusion} is devoted 
to a discussion of the results and to our final remarks. 
We assume metric signature $(-+++)$ and natural units in which
$c=G=\hbar=1$ unless stated otherwise.

\section{Awaking the vacuum of nonminimally coupled scalar
fields in the spacetime of rotating objects}
\label{sec:quantization}

Let us consider a spacetime which is nearly flat in
the asymptotic past and stationary and axially symmetric
in the asymptotic future corresponding to the formation 
of a rotating compact object from originally low density 
matter. In particular, let us assume that in the future 
the spacetime is well described by the line element 
\cite{Friedman2013}
\begin{equation} \label{general_metric}
	ds^2 = g_{00} dt^2 + 2 g_{03} dt d\varphi + g_{33} d\varphi^2
	+ g_{11} (dx^1)^2 + g_{22} (dx^2)^2,
\end{equation}
where $g_{\mu\nu} = g_{\mu\nu} (x^1, x^2)$ is assumed to be
smooth at the origin and continuous in the entire domain, 
while $(\partial_t)^\mu$ and $(\partial_\varphi)^\mu$ are 
(commuting) timelike and spacelike Killing fields, 
respectively.  Moreover, the whole spacetime 
is assumed to be asymptotically flat and bear no event or Cauchy 
horizons. 

In this fixed background, let us consider the massless 
Klein-Gordon equation
\begin{equation} \label{KG_eq}
	(- \nabla^\mu \nabla_\mu +\xi R) \phi =0
\end{equation}
describing the dynamics of a nonminimally coupled real scalar
field $\phi$, where $\xi \in \mathbb{R}$ and $R$ is the scalar 
curvature.

Now, let us restrict attention to the spacetime portion
described by the metric~(\ref{general_metric}) and
consider the following solution of Eq.~(\ref{KG_eq}) 
compatible with the spacetime symmetries and regular at the 
symmetry axis:
\begin{equation} \label{general_form_phi}
	\phi_{\omega m } (t, x^1, x^2, \varphi) 
	= e^{-i \omega t + i m \varphi}
	F_{\omega m }(x^1, x^2),
\end{equation}
where $\omega \in \mathbb{C}$, $m \in \mathbb{Z}$,
$F_{\omega m }=F_{\omega m }(x^1, x^2)$ satisfies the differential equation
\begin{align} \label{pde_F}
	& \frac{1}{\sqrt{-g}} \frac{\partial}{\partial x^p} \left(
	g^{pq} \sqrt{-g} \frac{\partial F_{\omega m }}{\partial x^q} \right)
	\nonumber \\
	+& \left( \frac{\omega^2 + m^2 g_{00}/g_{33} 
	+ 2 \omega m g_{03}/g_{33} }{- g_{00} + g_{03}^2/g_{33} }
	- \xi R \right) F_{\omega m } = 0
\end{align}
with $p,q \in \{1,2\}$ and $g \equiv \det(g_{\mu\nu})$.
Since the metric~(\ref{general_metric}) is assumed to be 
asymptotically flat, the behavior of $F_{\omega m }(x^1, x^2)$ 
at spatial infinity is given by 
\begin{equation} \label{asympt}
	F_{\omega m }(r, \theta) 
	\overset{r \to \infty}{\longrightarrow} 
	\sum_{\lambda=\pm} \sum_{l = |m|}^{\infty} {N_{\omega l m\lambda } P_l^m (\cos \theta ) 
	 \frac{e^{i \lambda \omega r}}{r}},
\end{equation}
where $N_{\omega l m \lambda} = {\rm const}$, $P_l^m (y)$ are associate Legendre 
polynomials, and we have chosen coordinates $\{x^1, x^2\}$ to reduce
asymptotically to the spherical ones $\{r, \theta\}$. The constants 
$N_{\omega l m \lambda}$ are determined by the regularity condition
at the symmetry axis up to an overall factor. 

For $\Im (\omega) \neq 0$ only solutions $F_{\omega m}$ with 
\begin{equation}
\lambda \Im(\omega) > 0
\label{condition}
\end{equation}
will be physically acceptable so that modes (\ref{general_form_phi}) 
are well behaved at spatial infinity. This constrains the acceptable values
of $\omega$ to those (if any) for which either $N_{\omega l m +} =0$
(in the case $\Im (\omega) <0$) or $N_{\omega l m -} =0$
(in the case $\Im (\omega) >0$):
\begin{equation} \label{davidstar}
	F_{\omega m }(r, \theta) 
	\overset{r \to \infty}{\longrightarrow} 
	\sum_{l = |m|}^{\infty} {N_{\omega l m \lambda} P_l^m (\cos \theta ) 
	\frac{e^{i \lambda \omega r}}{r}},
\end{equation}
where $\lambda \Im(\omega) > 0$. It should be 
noted that Eq.~(\ref{pde_F}) allows us to write the equality 
(up to an arbitrary multiplicative constant)
\begin{equation}
	F^*_{\omega m } (x^1, x^2) = 
	F_{\omega^* m } (x^1, x^2).
\label{identity}
\end{equation}

Here, we are interested in the case where unstable modes
of the form (\ref{general_form_phi}) exist. Classically, 
the existence of such unstable 
modes implies that generic linear perturbation out of the $\phi = 0$ 
equilibrium configuration grows unboundedly in time.
This unbounded amplification of classical linear perturbations 
indicates the breakdown of the test-field approximation, in which
the field evolves in a fixed background, and implies that the 
nonlinear interaction between the field and gravity must be 
taken into account. 

In Refs.~\cite{lv10,lmv10} it was shown that even in the
absence of classical perturbations quantum mechanics
provides a natural mechanism through which the instability
settles in by means of the amplification of quantum 
vacuum fluctuations and, consequently, of the field's
vacuum energy density. Indeed, in Ref.~\cite{mmv13}
we argue that if the initial mean field amplitude is much 
larger than $\sqrt{\hbar}$ a classical description of the
instability is suitable but if it is of the order of
$\sqrt{\hbar}$ then a quantum treatment should be employed. 
In what remains of this section, we discuss some aspects of 
the field quantization in the presence of unstable modes.
(See, e.g., Refs.~\cite{Unruh1974,Ford1975,Matacz1993} for
the quantization procedure in some stationary spacetimes 
for which the field is stable and Ref.~\cite{Lima2013} for 
a rigorous discussion on the quantization of unstable
fields in globally static spacetimes.)

In the canonical quantization procedure (see, e.g., 
Refs.~\cite{Fulling1989,Birrell1982}) the field and the associated 
momentum density are promoted to operators satisfying 
usual commutation relations. The field operator can
be expanded in a set of mode functions:
\begin{equation} \label{phi_expanded}
	\hat \phi = \int {d\mu (\sigma)
	[\hat{a}_{\sigma} u_{\sigma}^{(+)}
	+ \hat{a}_{\sigma}^\dagger u_{\sigma}^{(-)}]},
\end{equation}
where $d\mu(\sigma)$ is a measure on the set of quantum
numbers $\sigma$. The modes $u_{\sigma}^{(+)}$ and
$u_\sigma^{(-)} = (u_\sigma^{(+)})^*$ are positive- and 
negative-norm solutions of Eq.~(\ref{KG_eq}), respectively, satisfying 
\begin{equation} \label{KG_ip0}
(u_{\sigma}^{(\pm)}, u_{\sigma'}^{(\pm)})_{KG} = 
\pm \delta(\sigma, \sigma') 
\;\;\; {\rm and} \;\;\;
(u_{\sigma}^{(\pm)}, u_{\sigma'}^{(\mp)})_{KG} = 0,
\end{equation}
where the Klein-Gordon inner product $(,)_{KG}$
is defined by
\begin{equation} \label{KG_ip}
	(u,v)_{KG} \equiv i \int_{\Sigma_t}{d\Sigma 
	n^\mu [u^* \nabla_\mu v - v \nabla_\mu u^*]}
\end{equation}
with $\Sigma_t$ denoting a  Cauchy 
surface with proper volume element $d\Sigma$ and 
future-pointing unit normal 
vector field $n^\mu$.  The operator-valued
coefficients in Eq.~(\ref{phi_expanded}) satisfy
$[\hat{a}_{\sigma}, \hat{a}_{\sigma'}^\dagger]
= \delta(\sigma, \sigma')$
and zero for the remaining commutators. The vacuum $|0\rangle$ 
associated with this representation is then defined by requiring  
$\hat{a}_{\sigma} |0\rangle = 0$ for all $\sigma$.

Let us assume the quantum state 
to be the vacuum $|0_\textrm{in}\rangle$ defined with 
respect to a basis $\{  u_{\vec{k}}^{(\pm)} \}$ of modes 
which behave  as plane waves in the asymptotic past (where the 
spacetime is flat):
\begin{equation} \label{plane_wave}
	u_{\vec{k}}^{(\pm)} \overset{\textrm{past}}{\sim} 
	(16 \pi^3 |\vec{k}|)^{-1/2} 
	\exp[\mp i(|\vec{k}| t - \vec{k} \cdot \vec{x})],\;\;\; \vec{k} \in \mathbb{R}^3,
\end{equation} 
with $(t,\vec x)$ being usual Cartesian coordinates. Thus, 
$|0_\textrm{in}\rangle$ is the no-particle state according 
to static past observers.  

Let us now construct another set of orthonormal modes 
defined by their behavior in the asymptotic future. We
choose $\Sigma_t$ to be a $t = \textrm{const}$ hypersurface with
normal vector field $n^\mu =(1/N)(1, 0, 0, \Omega)$, where 
$N \equiv (-g_{00} + g_{03}^2/g_{33})^{1/2}$ and 
$\Omega \equiv - g_{03}/g_{33}$.
For this purpose, we first point out a useful property 
for solutions of Eq.~(\ref{pde_F}) with proper boundary
conditions. From Eq.~(\ref{pde_F}), we have
\begin{align} \label{trick}
	&\frac{\partial}{\partial x^p} \left[ 
	g^{pq} \sqrt{-g} \left( F_{\omega' m }
	\frac{\partial F^*_{\omega m }}{\partial x^q} - F^*_{\omega m }
	\frac{\partial F_{\omega' m }}{\partial x^q} \right)\right]
		\nonumber \\
	= &\frac{\sqrt{-g}}{N^2}(\omega' - \omega^*)
	 (\omega^* +\omega'- 2m \Omega) 
	F^*_{\omega m } F_{\omega' m }.
\end{align} 
Integrating Eq.~(\ref{trick}) by recalling 
Eqs.~(\ref{asympt})-(\ref{identity}), a nontrivial 
weighted orthonormality relation for $F_{\omega m}$
can be obtained:
\begin{equation} \label{norm_unstable}
	\int{dx^1 dx^2 \frac{\sqrt{-g}}{N^2} 
	(\omega + \omega'- 2 m \Omega)
	F_{\omega m } F_{\omega' m }}= 
	2\omega \delta_{\omega \omega'}
		\end{equation}
for
$\omega, \omega' \in \mathbb{C}-\mathbb{R}$ 
and
\begin{equation} \label{norm_stable}
	\int{dx^1 dx^2 \frac{\sqrt{-g}}{N^2} 
	(\omega + \omega' - 2 m \Omega)
	F_{\omega m }^* 	F_{\omega' m }}=2 \omega \delta(\omega - \omega') 
\end{equation}
for $\omega, \omega' \in \mathbb{R}$.

Now, we can construct a set of orthonormal solutions of 
Eq.~(\ref{KG_eq}) by determining their behavior at the 
asymptotic future. This set can in principle comprise both 
time-oscillatory (stationary) and tachyonic (non-stationary) 
modes. Positive-norm oscillatory modes read
\begin{equation} \label{u_pos}
	v_{\omega m}^{(+)} \overset{\textrm{future}}{\sim} 
	\frac{e^{-i\omega t + im \varphi}}{(4 \pi \omega)^{1/2}}
	 F_{\omega m} (x^1, x^2)
\end{equation}
with $\omega > 0$, while positive-norm tachyonic modes read
\begin{align} \label{w_pos}
	w_{\omega m}^{(+)} 
	& \overset{\textrm{future}}{\sim} 
	\sec(\alpha - \beta)^{1/2} \nonumber \\
	& \times \left[ (8 \pi \omega )^{-1/2}
	e^{-i \omega t + i m \varphi} e^{i \alpha} 
	F_{\omega m } (x^1, x^2) \right.\nonumber \\
	& + \left. (8 \pi \omega^*)^{-1/2} 
	e^{-i \omega^* t + i m \varphi}
	e^{i \beta} F^*_{\omega m } (x^1, x^2)\right]  
\end{align}
with $\Im(\omega)>0$ (see Eq.~(\ref{davidstar}); the principal 
square root is assumed). We note that by setting 
$\alpha = -\beta = \pi/6$, Eq.~(\ref{w_pos}) 
matches the form presented in Refs.~\cite{lv10,lmv10}
for the static case where $\omega$ is purely
imaginary. It can be verified that the set 
$\{v_{\omega m}^{(\pm)}, w_{\omega m}^{(\pm)}\}$
characterized by the asymptotic forms~(\ref{u_pos}) 
and~(\ref{w_pos}) is orthonormalized in agreement with 
Eq.~(\ref{KG_ip0}).

The existence of tachyonic modes (\ref{w_pos}) implies 
that at least some of the in-modes (\ref{plane_wave}) 
will go through a phase of exponential growth and, 
consequently, for a field in the in-vacuum state
$|0_\textrm{in}\rangle$, the expectation value of 
$\hat{\phi}^2$ will be exponentially amplified in time:
\begin{equation}
	\langle 0_\textrm{in}| \hat{\phi}^2 | 
	0_\textrm{in} \rangle \overset{\textrm{future}}{\sim}
	\frac{\hbar \kappa}{4 \pi |\bar{\omega}|}
	e^{2 \Im (\bar{\omega}) t}
	|F_{\bar{\omega} \bar{m} }|^2 
	[1 + O(e^{-\epsilon t})]
\end{equation}
(although 
$\langle 0_\textrm{in}| \hat{\phi} |0_\textrm{in} \rangle =0$).
Here, $\kappa\sim 1$ encodes information about the transition 
to the unstable phase, $\epsilon$ is some positive constant, 
$\bar{\omega}$ is the $\omega$ with largest value of 
$\Im(\omega)$ (achieved for a certain value of 
$m=\bar m$), and we have restored the $\hbar$.
This amplification of vacuum fluctuations leads to an 
exponential enhancement of the expectation value of the 
field's stress-energy-momentum tensor, as was discussed 
in Ref.~\cite{lv10}. The system then evolves according to
Einstein's semiclassical equations, at least while 
fluctuations of the field's stress-energy-momentum
tensor are relatively ``small'' \cite{Ford1993}.

In this paper, we will focus on searching for solutions in 
the form (\ref{general_form_phi}) with $\Im(\omega)>0$, which 
are regular at the origin and vanish at spatial infinity 
[see Eq.~(\ref{davidstar}), where $\lambda=+$]. 
Normalized tachyonic modes can be constructed from these
solutions by adjusting the normalization as in 
Eq.~(\ref{norm_unstable}). Our main purpose will be 
to understand how the range of field couplings $\xi$ 
for which unstable modes appear changes due to rotation.

\section{First order deviations from staticity}
\label{sec:1storder}

Here, we argue that in order to extract nontrivial results
concerning the instability analysis, we must go beyond 
first order deviations from staticity. First, let us assume 
that the metric components in Eq.~(\ref{general_metric}) are 
analytic functions of $J/M^2$ so that a perturbative 
treatment  for small $J/M^2$ is meaningful, where $M$ and $J$ 
are mass and angular momentum of the compact object 
(computed, e.g., by Komar formulas). 

Physically, it is clear that the field instability cannot 
depend on the rotation direction. Mathematically, this can be seen
as follows. First, we note that rotation reversal, $J \to -J$,
is equivalent to time reversal, $t \to -t$. 
This implies that $g_{03}$ is an odd function of $J$, while the 
remaining metric components, as well as $R$, are even. Then,
for every regular solution $F_{\omega m }^{(\xi, J,M,\ldots)}$
of Eq.~(\ref{pde_F}) there will exist a corresponding one:
\begin{equation*}
	F_{\omega - m }^{(\xi, -J, M, \ldots )}
	\propto 
	F_{ \omega m }^{( \xi, J, M, \ldots )},
\end{equation*}
where $(\xi, J, M,  \ldots )$ was added to explicitly label  all field
and spacetime parameters on which $F_{\omega  m}$ depends. Since modes 
with all values of $m$ enter in the field expansion, we conclude that 
whenever we have instability for a configuration 
$(\xi, J, M, \ldots)$ the same will be true for $(\xi, -J, M, \ldots)$. 
In particular,  
$$
\xi_0(J, M, \ldots) = \xi_0( -J, M, \ldots),
$$   
where
$\xi_0=\xi_0 (J, M, \ldots)$ is the value of $\xi$ which marks 
the appearance of (any) tachyonic modes as a function of the spacetime 
parameters. As a result, in order to see effects due to rotation in 
$\xi_0$, we must carry out our expansion at least up to second order 
in $J/M^2$.

Much less intuitive is the fact that   
$$
\xi_{0; \;m, \Im(\omega)} (J, M, \ldots) = \xi_{0; \;m, \Im(\omega)} (-J, M, \ldots),
$$
where $\xi_{0; \; m, \Im(\omega)}=\xi_{0; \; m, \Im(\omega)} (J, M, \ldots)$
is the value of $\xi$ which marks the appearance of a tachyonic mode with 
quantum numbers $m$ and $\Im(\omega)$ (we have omitted $\Re(\omega)$, since 
it is irrelevant for the instability). The fact that 
$\xi_{0; \;m, \Im(\omega)}$ is an even function of $J/M^2$ can be traced back 
to the fact that if $F_{\omega m}^{(\xi, J, M, \ldots)}$ 
is a regular solution of Eq.~(\ref{pde_F}), the same is true for 
${F_{-\omega^* m}^{(\xi, -J, M,\ldots)}}$, since
\begin{equation*}
    {F_{\omega \,m }^{(\xi, J, M, \ldots)}}^*
	\propto 
	{F_{-\omega^* \,-m }^{(\xi, J, M, \ldots )}}
	\propto 
	{F_{-\omega^* \,m }^{(\xi, -J, M, \ldots )}}.
\end{equation*}

From the discussion above, we conclude that corrections due to rotation
to $\xi_0$ and $\xi_{0; \;m, \Im(\omega)}$
are of even order on the parameter $J/M^2$ (which will be 
manifest in the results of Sec.~\ref{sec:2ndorder}). Thus, in what 
follows, we will explore second-order corrections in a particular 
spacetime, which we now describe.

\section{Rotating thin shells}
\label{sec:spacetime}

The Kerr metric, given in Boyer-Lindquist coordinates by
\begin{align} \label{kerr}
	ds^2 
	& = -\left( 1 - \frac{2Mr}{r^2 + a^2 \cos^2 \theta} \right)
	dt^2 - \frac{4 a M r \sin^2 \theta}{r^2 + a^2 \cos^2\theta}
	dt d\varphi \nonumber \\
	& + (r^2 + a^2 \cos^2 \theta) 
	\left( \frac{dr^2}{r^2 - 2 M r + a^2} 
	+ d \theta^2 \right) \nonumber \\
	&+ \left( r^2 + a^2 + 
	\frac{2 M r a^2 \sin^2 \theta}{r^2 + a^2 \cos^2 \theta}
	\right) \sin^2 \theta d\varphi^2,
\end{align}
besides being the only vacuum solution of Einstein's equations 
describing stationary black holes, can in principle also
approximate the gravitational field outside an axially symmetric
rotating source with mass $M$ and angular momentum 
$J=aM$. 
In particular, in Ref.~\cite{Cruz1968} a spinning shell was
considered as a source of the Kerr metric and the matching 
of internal and external solutions was worked out explicitly 
up to third order in the rotation parameter. In this section, 
we will describe in some detail the particular case of a flat 
interior matched with an external Kerr field up to second order 
in $a/M$, which will suffice as a prototype model of a 
rotating system.

Therefore, let us consider a stationary and axially symmetric
thin shell of matter surrounded by vacuum. The spacetime region
internal to the shell is taken to be flat, with line element
\begin{equation} \label{interior}
	ds^2_- = - d\tau^2 + d\rho^2 + \rho^2 
	(d\Theta^2 + \sin^2 \Theta d\Phi^2 ),
\end{equation}
while the external-to-the-shell portion of the spacetime 
will be described by the Kerr metric, Eq.~(\ref{kerr}), 
expanded up to second order in $a/M$:
\begin{align} \label{kerrapprox}
	ds^2_+ 
	&= - \left [1 - \frac{2M}{r} \left(1-\frac{a^2}{r^2}
	\cos^2\theta \right) \right] dt^2 
	- \frac{4 a M}{r} \sin^2 \theta dt d\varphi \nonumber \\
	&+ \left( r^2 + a^2 \cos^2 \theta 
	- \frac{a^2 r^2}{r^2 - 2 M r} \right)
	\frac{dr^2}{r^2 - 2 M r} \nonumber \\
	& + \left[ r^2 + a^2 
	\left( 1 + \frac{2 M}{r} \sin^2 \theta \right) \right]
	\sin^2 \theta d\varphi^2 \nonumber \\
	& +	(r^2 + a^2 \cos^2 \theta) d\theta^2. 
\end{align}
This approximation is valid as long as the corresponding 
error is small, i.e., 
$g_{\mu\nu}^\textrm{kerr} - g_{\mu\nu}^\textrm{approx} 
\ll g_{\mu\nu}^\textrm{kerr}$, which is satisfied if
\begin{equation} \label{validity}
	a^2 \ll M^2 \qquad \textrm{and} \qquad r \gg 2M.
\end{equation}
The label ``$-$'' (``$+$'') is used above and in what follows 
to indicate the restriction of certain quantities to the inner 
(outer) spacetime region with respect to the shell's worldtube, 
which we denote by $\su$. 

Equations (\ref{interior}) and~(\ref{kerrapprox}) will
represent portions of a single spacetime (with a singular
three-dimensional timelike boundary $\su$ between them)
provided that the internal and external metrics induced 
on $\su$, denoted by $h_{ab}$, coincide. Indeed, in 
Ref.~\cite{Cruz1968}, this was shown to be possible 
if $\su$ is determined by
\begin{equation} \label{rS}
	r|_\su = r_\su (\theta) \equiv \ra 
	\left[ 1 - \frac{a^2 \mathsf{F}^2}{\ra^2}  \cos^2 \theta \right],
\end{equation}
where $\ra= {\rm const}>0$ is the shell equatorial 
radial coordinate and 
\begin{equation} \label{F}
	\mathsf{F} \equiv \sqrt{1 - 2 M/\ra}.
\end{equation}
Note that since Eq.~(\ref{kerrapprox}) 
is only reliable in the regime given by
conditions~(\ref{validity}), it is necessary that at least
$	\ra > 2M$.
It is convenient to cover $\su$ with coordinates 
$\zeta^a = (t,\theta,\varphi)$, $a = 0,2,3$, since the 
shell lies at $r = r_\su (\theta)$ [see Eq.~(\ref{rS})].
Then, the continuity condition above allows 
us to relate the internal coordinates on $\su$ with $\zeta^a$ as
\begin{equation} \label{matching_cond}
	\left. \tau \right|_\su = A t, \;\;\;
	\left. \rho \right|_\su = \rho_\su (\theta), \;\;\;
	\left. \Theta \right|_\su = \Theta_\su (\theta), \;\;\;
	\left. \Phi \right|_\su = \varphi - \tilde{\Omega} t,
\end{equation}
where
\begin{align}
	A & \equiv \mathsf{F} 
	\left(1 + \frac{2 a^2 M^2}{\ra^4\mathsf{F}^2}\right), 
	\label{tauS} \\
	\rho_\su(\theta) &\equiv \ra \left[ 1 + \frac{a^2}{2 \ra^2}
	\left(1 + \frac{2M}{\ra} - 3 \cos^2 \theta \right) \right], 
	\label{rhoS} \\	
	\Theta_\su (\theta) &\equiv \theta + \frac{a^2}{2 \ra^2} 
	\left( 1 + \frac{2M}{\ra}\right) \sin\theta \cos\theta, 
	\label{ThetaS} \\
	\tilde{\Omega} & \equiv \frac{2 a M}{\ra^3}. 
	\label{Omega}
\end{align}

The shell can be shown to be slightly oblate according to 
zero-angular-momentum observers, since on a 
$t={\rm const}$ section of $\su$, 
\begin{equation} \label{Leq_per_Lmer}
	\frac{L_\textrm{equatorial}}{L_\textrm{meridional}} 
	= 1 + \frac{3 a^2}{4 \ra^2} \geq 1,
\end{equation}
where $L_\textrm{equatorial}$ and $L_\textrm{meridional}$ 
are the equatorial ($\theta = \pi/2$) and meridional 
($\varphi = \textrm{const}$) shell proper lengths, respectively. 
Note also that in this approximation the shell rotates 
rigidly with angular velocity \cite{Cruz1968}
\begin{equation*}
	\Omega_\textrm{shell} = d\varphi/ dt = 
	\tilde{\Omega} (1 + 2\mathsf{F}) 
	(1 - \mathsf{F})^{-1} (1 + 3 \mathsf{F})^{-1}
\end{equation*}
as measured by static observers at infinity, where
$\tilde{\Omega}$ was defined in Eq.~(\ref{Omega}). 

Once the spacetime is determined, the 
stress-energy-momentum tensor of the corresponding matter layer
is also fixed (see, e.g., Ref.~\cite{Poisson2004}):
\begin{equation} \label{Tmunu}
	T^{\mu\nu} = S^{ab} e^\mu_a e^\nu_b \delta (\ell ),
\end{equation}
where $\ell$ is the proper distance along geodesics which
intercept $\su$ orthogonally (such that $\ell<0$, $\ell=0$, 
and $\ell>0$ inside, on, and outside $\su$, respectively), 
$e^\mu_a \equiv \partial x^\mu/\partial \zeta^a$ are 
the components of the coordinate vectors 
$\partial/\partial \zeta^a = ( \partial_t, \partial_\theta,
\partial_\varphi)$ defined on $\su$, and 
\begin{equation} \label{Sab}
	S^{ab} = -\frac{1}{8 \pi}
	(\Delta K^{ab} - h^{ab} \Delta K)
\end{equation}
is the surface stress-energy-momentum tensor of the shell. 
Here, $K_{ab}$ is the extrinsic curvature, 
$K \equiv K_{ab} h^{ab}$, and 
$\Delta A^{a b c \ldots}_{\;\;\;m n o \ldots}$ denotes the
discontinuity of some quantity 
$A^{a b c \ldots}_{\;\;\; m n o \ldots}$ across $\cal{S}$. 
A direct calculation, following Ref.~\cite{Cruz1968}, leads to
\begin{align}
	8 \pi S_0^{\;\; 0} 
	& = \frac{2 \mathsf{F}}{\ra} - \frac{2}{\ra} 
	- \frac{a^2}{\ra^3} \left( 2 - 2\mathsf{F} + 
	\frac{M}{\mathsf{F}\ra} - \frac{2M}{\ra} \right)
	\nonumber \\
	& - \frac{3 a^2}{\ra^3} \left( 2\mathsf{F} 
	- 2 + \frac{3 M \mathsf{F}}{2 \ra} 
	- \frac{M}{2 \mathsf{F} \ra} 
	- \frac{M^2}{\mathsf{F} \ra^2} \right) \cos^2 \theta,
\end{align}
\begin{align}
	8\pi S_2^{\;\; 2} 
	& = \frac{\mathsf{F}}{\ra} + \frac{M}{\mathsf{F} \ra^2} 
	- \frac{1}{\ra} + \frac{a^2}{2 \ra^3} 
	\left( 1 - \frac{1}{\mathsf{F}} + \frac{2M}{\ra} 
	- \frac{M}{\ra \mathsf{F}^3} \right)
	\nonumber \\
	& + \frac{3 a^2}{2 \ra^3} \left( 1 - \mathsf{F} 
	- \frac{M \mathsf{F}}{\ra} \right) \cos^2\theta,
\end{align}
\begin{align}
	8\pi S_3^{\;\; 3} 
	& = \frac{\mathsf{F}}{\ra} + \frac{M}{\mathsf{F} \ra^2} 
	- \frac{1}{\ra} + \frac{a^2}{\ra^3} \left( 2\mathsf{F} 
	+ \frac{1}{2 \mathsf{F}} - \frac{5}{2} 
	+\frac{M}{\ra} \right.
	\nonumber \\
	& \left. + \frac{M \mathsf{F}}{\ra} 
	+ \frac{2 M^2}{\mathsf{F}\ra^2} 
	- \frac{M}{2 \ra \mathsf{F}^3} \right)
	+ \frac{a^2}{\ra^3} \left( \frac{9}{2} 
	- \frac{9 \mathsf{F}}{2} -\frac{2 M}{\mathsf{F} \ra} 
	\right. 
	\nonumber \\
	& \left. - \frac{5 M \mathsf{F}}{2 \ra} 
	- \frac{2 M^2}{\mathsf{F} \ra^2} \right) \cos^2\theta,
\end{align}
\begin{equation}
	8\pi S_0^{\;\; 3} 
	= - \frac{a M}{\ra^4} \left( 2 
	+ \frac{1}{\mathsf{F}} \right),
	\qquad 
	8\pi S_3^{\;\; 0} 
	= \frac{3 M a}{\mathsf{F} \ra^2} \sin^2\theta.
\end{equation}
A result that will be particularly useful later is
\begin{align} \label{DK}
	\Delta K 
	&= \frac{2 \mathsf{F}}{\ra} + \frac{M}{\mathsf{F} \ra^2} 
	- \frac{2}{\ra} + \frac{a^2}{\ra^3}
	\left( 2 \mathsf{F} - 2 + \frac{2M}{\ra} 
	- \frac{M}{2 \ra \mathsf{F}^3} \right) 
	\nonumber \\
	& + \frac{3a^2}{\ra^3 } \left( 2 - 2\mathsf{F} 
	- \frac{3 M \mathsf{F}}{2 \ra} \right) \cos^2 \theta.
\end{align}

It can be verified from the stress-energy-momentum tensor 
written above that the shell gravitational mass and angular
momentum are indeed equal to $M$ and $aM$,
respectively. Other physical properties can be more readily
investigated through the eigenvalues of $S_{ab}$. For this
purpose, let us define vectors $\hat{e}^a_\alpha$ that form 
an orthonormal basis on $\su$, in the sense that
\begin{equation*}
	h_{ab} \hat{e}^a_\alpha \hat{e}^b_\beta 
	= \eta_{\alpha\beta},
\end{equation*}
where $\eta_{\alpha \beta} = \textrm{diag}(-1,1,1)$. 
The surface density and pressures are then eigenvalues with
respect to these vectors and can be computed to be
\begin{align}
	\sigma & = S_{ab} \hat{e}^a_0 \hat{e}^b_0 
	= - S_0^{\;\; 0} + \frac{S_0^{\;\; 3} S_3^{\;\; 0}}
	{ (S_3^{\;\; 3})^{(0)} - (S_0^{\;\; 0})^{(0)}}, \\
	p_\theta & = S_{ab} \hat{e}^a_2 \hat{e}^b_2 
	= S_2^{\;\; 2}, \\
	p_\varphi & = S_{ab} \hat{e}^a_3 \hat{e}^b_3 = S_3^{\;\; 3} 
	+ \frac{S_0^{\;\; 3} S_3^{\;\; 0}}
	{(S_3^{\;\; 3})^{(0)} - (S_0^{\;\; 0})^{(0)}},
\end{align}
where we have introduced the superscript ``${(0)}$" in some terms to 
indicate that only the contributions of zeroth order 
in $a/M$ need to be considered to keep the
approximation consistent up to second order.

\begin{figure}[t]
\includegraphics[width=8.5cm]{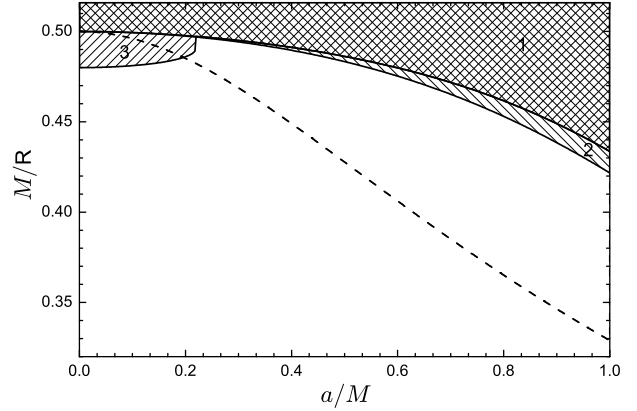}
\caption{The shaded region indicates the values of $M/\ra$ 
which are excluded by the weak, strong or dominant energy 
conditions, in terms of the rotation parameter $a/M$. 
The weak, strong, and dominant energy conditions are not satisfied
in region 1, in the union of regions 1 and 2, and in the union 
of regions 1 and 3, respectively. When $a=0$, we get the static 
limit, in which case the dominant energy condition is violated for 
$M/\ra > 0.48$. The region below the dashed line corresponds to 
values of $M/\ra$ where the error in approximating Eq.~(\ref{kerr}) 
by Eq.~(\ref{kerrapprox}) is less than $10\%$.}
\label{fig:EC}
\end{figure}

The classical energy conditions can be explicitly stated 
in terms of these eigenvalues (see, e.g., 
Ref.~\cite{Poisson2004}). The white region in Fig.~\ref{fig:EC} 
shows the values of $M/\ra$ for which the weak,
strong, and dominant energy conditions are satisfied as 
a function of $a/M$. The region below the dashed line indicates 
the values of $M/\ra$ which comply with the condition 
\begin{equation} \label{criterion}
	( g_{\mu\nu}^\textrm{kerr} - g_{\mu\nu}^\textrm{approx})
	/ g_{\mu\nu}^\textrm{kerr} < 0.1.
\end{equation}
In the subsequent analysis, we will only consider 
shells with $M/\ra$ below this dashed line. 

\section{Second order deviations from staticity in a shell model}
\label{sec:2ndorder}

In this section, we aim at investigating the influence of
rotation on the parameter space that characterizes unstable
configurations. We take as a model of a rotating system
the slowly spinning shells described in Sec.~\ref{sec:spacetime}
and look for unstable solutions of Eq.~(\ref{KG_eq}) 
in this spacetime. 

Let $\phi^-$ and $\phi^+$ 
denote solutions of $\nabla_\mu \nabla^\mu \phi^\pm=0$ in the 
inner and outer spacetime portions with respect to the shell's
worldtube~$\su$, respectively. They combine to form a solution of Eq.~(\ref{KG_eq})
in the entire spacetime provided that they are continuous at $\su$:
\begin{equation} \label{continuity}
	\left. \phi^- (\tau, \rho, \Theta, \Phi ) \right|_\su = 
	\left. \phi^+ (t, r, \theta, \varphi ) \right|_\su
\end{equation}
and their derivative along the direction orthogonal to 
the shell is discontinuous by a definite amount:
\begin{equation} \label{discontinuity}
	\left. \Delta (d\phi / d\ell ) \right|_\su 
	= \left. - 2 \xi \; \Delta K \phi \right|_\su,
\end{equation}
which follows from Eq.~(\ref{KG_eq}) if we notice from 
Eq.~(\ref{Tmunu}) that
\begin{equation*} 
	R = - 8 \pi T = -2 \Delta K \delta ( \ell ),
\end{equation*}
where $\Delta K$ is given in Eq.~(\ref{DK}). Equations 
(\ref{continuity}) and (\ref{discontinuity}), together with
appropriate boundary conditions on $\phi^\pm$, determine
uniquely the joined solution.

In the interior of $\su$, solutions of the form~(\ref{general_form_phi})
can be written as
\begin{equation} \label{v_in_2}
	\phi^-_{\omega' m'} (\tau, \rho, \Theta, \Phi)
	= \sum_{l = |m'| }^\infty 
	{ N^-_{ \omega' l m'} e^{- i \omega' \tau}
	\chi_{\omega' l}(\rho )}
	Y_{l m'} (\Theta, \Phi ),
\end{equation}
where $\omega' \in \mathbb{C}$, $m' \in \mathbb{Z}$,
$Y_{l m'}(\Theta, \Phi )$ are spherical harmonics, 
$N^-_{\omega' l m'}$ are arbitrary (complex) constants
and $\chi_{\omega' l}(\rho )$ satisfy the spherical Bessel 
equation,
\begin{equation} \label{bessel_eq}
	\rho^2 \frac{d^2 \chi_{\omega' l}}{d \rho^2} 
	+ 2 \rho \frac{d \chi_{\omega' l}}{d \rho}
	+[ \omega'^2 \rho^2 - l (l + 1)] \chi_{\omega' l} = 0,
\end{equation}
with the additional condition of regularity at the origin, 
so that, for $\omega' \neq 0$, $\chi_{\omega' l}(\rho) 
\propto j_l (\omega' \rho)$.
A summation is included in Eq.~(\ref{v_in_2}), since the 
spacetime is not spherically symmetric and, as we will
show below, the matching in Eq.~(\ref{continuity}) will
mix different values of $l$.

In the region external to $\su$, we analogously write
\begin{equation} \label{v_out}
	\phi^+_{\omega m} (t, r, \theta, \varphi )
	= \sum_{l = |m|}^{\infty} {N^+_{\omega l m} 
	e^{- i \omega t + i m \varphi}
	\psi_{\omega l m} (r) S_{\omega l m} (\cos \theta )}
\end{equation}
with $\omega \in \mathbb{C}$ and $m \in \mathbb{Z}$.
Here, $S_{\omega lm} (\cos\theta)$ are spheroidal 
harmonics~\cite{Flammer1957} satisfying
\begin{align}
& \frac{1}{\sin \theta} \frac{d}{d\theta}
\left( \sin \theta \frac{d S_{\omega lm}}{d\theta} \right)
\nonumber \\
&+\left( 
\Lambda_{lm}+a^2 \omega^2 \cos^2 \theta - \frac{m^2}{\sin^2 \theta} 
\right)
S_{\omega lm} = 0,
\end{align}
where
$$
\Lambda_{l m} = l(l+1)- \frac{(2l^2+2l-2m^2-1)a^2\omega^2}{(2l-1)(2l+3)} 
+ O(a^4 \omega^4 ).
$$
The expansion of $S_{\omega lm}(y)$ in powers of the
dimensionless parameter $a \omega$ has the following leading terms:
\begin{align*}
	S_{\omega l m}(y) 
	& = P_l^{m}(y) + a^2 \omega^2 
	\left[ - \frac{(l + m - 1) (l + m)}{2 (2 l + 1)(2 l - 1)^2}
	P_{l - 2}^m (y) \right. \nonumber \\
	&\left. + \frac{(l - m + 1)(l - m + 2)}{2 (2 l + 1)
	(2 l + 3)^2} P_{l + 2}^m (y) \right] + O(a^4 \omega^4 ).
\end{align*}
The radial functions $\psi_{\omega l m} (r)$ obey the 
differential equation (up to second order)
\begin{align} \label{psi2_eq}
	& f \left( f + \frac{2 a^2}{r^2} \right) 
	\frac{ d^2 \psi_{\omega l m}}{dr^2} 
	+ \frac{2}{r} \left( f + \frac{a^2}{r^2} \right) 
	\left( 1 - \frac{M}{r} \right)
	\frac{d \psi_{\omega l m}}{dr} 
	\nonumber \\
	& + \left[ \omega^2 \left( 1 + \frac{a^2}{r^2} 
	+ \frac{2 M a^2}{r^3} \right)
	- \frac{4 a M \omega m}{r^3} + \frac{m^2 a^2}{r^4}
	\right. 
	\nonumber \\
	& -\left( 
	\left.
	\frac{\Lambda_{lm} f}{r^2} + \frac{l(l+1) a^2}{r^4}
	\right) 
	\right] \psi_{\omega l m} = 0,
\end{align}
where $f = f(r) \equiv 1 - 2 M / r$. 
Since we are looking for normalizable solutions with 
$\Im(\omega)>0$, $\psi_{\omega lm} (r)$ must go
asymptotically as [see Eq.~(\ref{davidstar})]
\begin{equation} \label{decay}
	\psi_{\omega l m} (r) \stackrel{r \to + \infty}{\propto} 
	\frac{e^{i \omega r}}{r}.
\end{equation}

The continuity condition, Eq.~(\ref{continuity}), implies
\begin{equation} \label{mu_m}
	m' = m,
\end{equation}
\begin{equation} \label{varpi_omega}
	\omega' 
	= \frac{\omega - m \tilde{\Omega}}{A} 
	= \frac{1}{\mathsf{F}}
	\left( \omega - \frac{2 M m a}{\ra^3}
	- \frac{2 \omega a^2 M^2}{\ra^4 \mathsf{F}^2} \right),
\end{equation}
and
\begin{align} \label{cont2}
	& \sum_{l\geq |m|} N^+_{\omega l m} S_{\omega l m}(\cos\theta )
	\psi_{\omega l m} [r_\su (\theta) ]
	\nonumber \\
	& = \sum_{l\geq |m|} {N^-_{\omega' l m}
	P_l^m [\cos\Theta_\su (\theta) ] \chi_{\omega' l} 
	[\rho_\su (\theta)]},
\end{align}
where we recall that $\mathsf{F} = \sqrt{f(\ra)}$ 
[see Eq.~(\ref{F})] and that $r_\su(\theta)$, 
$\rho_\su(\theta)$, and $\Theta_\su(\theta)$ are defined 
in Eqs.~(\ref{rS}), (\ref{rhoS}), and~(\ref{ThetaS}), 
respectively. Equation~(\ref{mu_m}) comes from the spacetime
axial symmetry, while Eq.~(\ref{varpi_omega}) relates the
energies ascribed to a certain mode by an  inertial 
observer inside the shell with proper time $\tau$ and by a static
observer at spatial infinity: $\omega$ and $\omega'$ differ not only due 
to redshift but also due to the coupling between 
rotation and the mode's angular momentum.

Equation (\ref{cont2}) can be used to express the coefficients 
$N^+_{\omega l m}$ in terms of $N^-_{\omega' l m}$. For this
purpose, it will be useful to Taylor expand 
$\chi_{\omega' l} [\rho_\su (\theta) ]$ 
and $\psi_{\omega l m} [r_\su (\theta) ]$
around $\rho = \ra$ and $r = \ra$, respectively [see Eqs.~(\ref{rhoS})
and~(\ref{rS})],
\begin{equation} \label{chi_norm}
	\chi_{\omega' l} [\rho_\su (\theta) ] \! =\! 
	\chi_{\omega' l}(\ra)+ \frac{a^2}{2\ra} \left( 1 + \frac{2M}{\ra}
	- 3 \cos^2 \theta \right)
	\left. \! \frac{d \chi^{(0)}_{\omega' l}(\rho)}{d\rho}
	\right|_{\ra},
\end{equation}
\begin{equation} \label{psi_norm}
	\psi_{\omega l m} [r_\su (\theta) ]
	= \psi_{\omega lm} (\ra)- \frac{a^2}{\ra} \mathsf{F}^2 \cos^2\theta 
	\left. \frac{d\psi^{(0)}_{\omega l m}(r)}{dr}
	\right|_{\ra},
\end{equation}
and to fix  
$\chi_{\omega' l}(\ra) = 1$ and $\psi_{\omega lm} (\ra) = 1$,
which can be done with no loss of generality. 
The arbitrariness in the
normalization of $\phi^-_{\omega'm}$ and $\phi^+_{\omega m}$ 
will be completely encoded in
$N^-_{\omega' l m}$ and $N^+_{\omega l m}$, which can be 
adjusted in order to comply with Eq.~(\ref{norm_unstable}).
Then, by writing 
\begin{equation} \label{N-}
	N_{\omega' l m}^- = N_{\omega' l m}^{- (0)} 
	+ \frac{a^2}{M^2} N_{\omega' lm}^{- (2)}
\end{equation}
and similarly for $	N_{\omega l m}^+$,
we note that for $a=0$ Eq.~(\ref{cont2}) implies 
$N_{\omega lm}^{+ (0)} = N_{\omega' lm}^{- (0)}$. 
Therefore, up to $O(a^2/M^2)$, we can write 
\begin{equation} \label{ansatz}
	N_{\omega l m}^+ = N_{\omega' l m}^- + 
	\frac{a^2}{M^2} g_{\omega' l m}, \;\;\;l \geq |m|.
\end{equation}
Inserting Eq.~(\ref{ansatz}) and 
Eqs.~(\ref{chi_norm})-(\ref{psi_norm}) 
in Eq.~(\ref{cont2}), multiplying the
latter by $P_{l'}^m (\cos\theta)$ and integrating over 
$\theta$, we obtain, after some algebra,
\begin{equation} \label{rel_coef}
	g_{\omega' l m} = N^{- (0)}_{\omega' l m}
	\alpha_1 + N^{- (0)}_{\omega' (l + 2) m} \alpha_2 
	+ N^{- (0)}_{\omega' (l - 2) m} \alpha_3 \; H(l - |m| - 2),
\end{equation}
where $H(x)$ is the Heaviside step function,
\begin{align*}
	\alpha_1 
	& = \frac{M^2}{\ra} \left( \frac{M}{\ra} 
	- \frac{ l^2 + l - 3m^2}{(2 l + 3)(2 l - 1)}\right)
	\left. \frac{d \chi^{(0)}_{\omega' l}(\rho)}{d\rho}
	\right|_{\ra} 
	\nonumber \\
	& + \frac{M^2 \mathsf{F}^2}{\ra} 
	\left(
	\frac{2 l^2 + 2 l - 1 - 2 m^2}{(2 l + 3)(2l - 1)}
	\right)
	\left. \frac{ d\psi^{(0)}_{\omega l m}(r)}{dr}
	\right|_{\ra}
	\nonumber \\
	& -\frac{M^2}{2 \ra^2} 
	\left(\frac{l^2 + l - 3m^2}{(2l + 3)(2l - 1)}\right) 
	 \left( 1 + \frac{2M}{\ra} \right),
\end{align*}
\begin{align*}
	\alpha_2 
	&= \frac{M^2}{\ra}
	\left(\frac{(l + m + 1)(l + m + 2)}{(2l + 3)(2l + 5)}\right)
	\left[ 
	- \frac{l + 3}{2\ra} \left( 1 + \frac{2M}{\ra} \right)  \right.
	\nonumber \\
	& \frac{\omega^2 \ra}{4 l + 6} 
	+ \left. \mathsf{F}^2 \left. 
	\frac{d\psi^{(0)}_{\omega (l + 2) m}(r)}{dr} \right|_{\ra}
	- \frac{3}{2} \left. 
	\frac{d \chi^{(0)}_{\omega' (l + 2)}(\rho)}{d\rho}
	\right|_{\ra} \right],
\end{align*}
and
\begin{align*}
	\alpha_3 
	& = \frac{M^2}{\ra}
	\left( \frac{(l - m - 1)(l - m)}{(2l - 1)(2l - 3)} \right)
	\left[ 
	\frac{l - 2}{2\ra} \left( 1 + \frac{2M}{\ra}\right) \right.
	\nonumber \\
	& - \left. \frac{\omega^2 \ra}{4l - 2} 
	+ \mathsf{F}^2
	\left. \frac{d\psi^{(0)}_{\omega (l - 2) m}(r)}{dr}
	\right|_{\ra} 
	- \frac{3}{2}
	\left. \frac{d \chi^{(0)}_{\omega' (l - 2)}(\rho)}{d\rho}
	\right|_{\ra} \right].
\end{align*}
It is worthwhile to note in Eq.~(\ref{rel_coef}) the coupling 
between multipolar indices $l$ and $l \pm 2$ that appears in 
$O(a^2/M^2)$ due to the absence of spherical symmetry 
[see Eq.~(\ref{Leq_per_Lmer})]. More generic deviations 
from spherical symmetry, such as those considered in 
Ref.~\cite{lmmv13}, can give rise to a more involved mixing.

\begin{figure}[t]
\includegraphics[width=8cm]{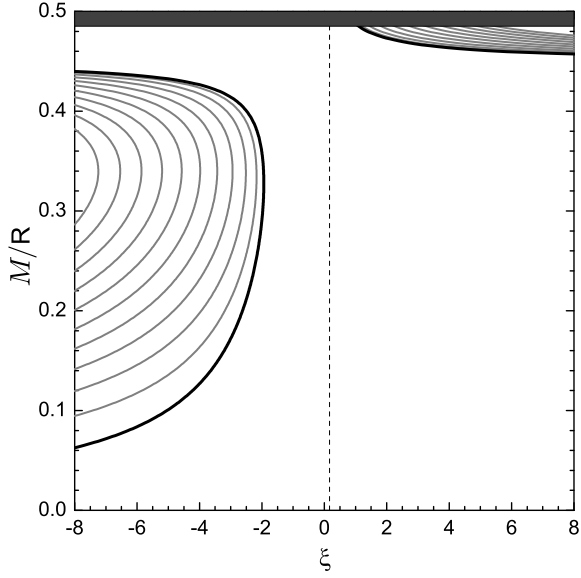}
\caption{Diagram showing the regions in the 
parameter space $(\xi,M/\ra)$ for which the instability 
is triggered when $a/M=0.2$. 
The black curves are characterized by $\omega_I \approx 0$ 
and provide the boundary of the unstable regions. 
Internal gray curves are characterized by values of $\omega_I$ which 
increase in steps of $0.04$ up to $\omega_I=0.4$.
Here, we have set $l_0=m=0$. The black strip excludes values
of $M/\ra$ for which Eq.~(\ref{criterion}) does not hold
and the vertical dashed line indicates the
conformal-coupling value $\xi = 1/6$.}
\label{fig:omega_var}
\end{figure}

\begin{widetext}
The discontinuity condition on the derivatives, 
Eq.~(\ref{discontinuity}), can be more explicitly written as
\begin{align} \label{discont2}
	& \sum_{l\ \geq |m|} { N^+_{\omega l m}}
	\left[ \left. \frac{dr}{d\ell} \right|_\su
	\left.\frac{d\psi_{\omega l m} (r)}{dr} \right|_{r_\su} 
	S_{\omega l m} (\cos\theta) 
	+ \left. \frac{d\theta}{d\ell} \right|_\su
	\frac{d S_{\omega l m} (\cos \theta)}{d\theta}
	\psi_{\omega l m} (r_\su) \right]
	\nonumber \\
	& - \sum_{l\ \geq |m|} { N^-_{\omega' l m}}
	\left[ \left. \frac{d\rho}{d\ell} \right|_\su
	\left. \frac{d\chi_{\omega' l} (\rho)}{d\rho}
	\right|_{\rho_\su} P_l^m (\cos \Theta_\su )
	+\frac{d\Theta}{d\ell} \right|_\su
	\left. \frac{d P_l^m (\cos\Theta)}{d\Theta} 
	\right|_{\Theta_\su}
	\chi_{\omega' l}(\rho_\su) \bigg ] 
	\nonumber \\
	& = - 2 \xi \Delta K \sum_{l\ \geq |m|} {N^-_{\omega' l m} 
	\chi_{\omega' l}(\rho_\su)
	P_l^m(\cos\Theta_\su)},
\end{align}
\end{widetext}
where the $\theta$-dependence has been omitted in several 
terms and
\begin{equation*}
	\left. \frac{dr}{d\ell} \right|_\su 
	= \mathsf{F} - \frac{a^2 \mathsf{F}}{2 \ra^2}
	\left( 1 + \frac{2M}{\ra} \right) \cos^2\theta 
	+ \frac{a^2}{2 \ra^2 \mathsf{F}},
	\qquad \left.\frac{d\rho}{d\ell} \right|_\su = 1,
\end{equation*}
\begin{equation*}
	\left. \frac{d\theta}{d\ell}\right|_\su 
	= - \frac{2 a^2 \mathsf{F}}{\ra^3}
	\sin\theta \cos\theta, 
	\qquad
	\left. \frac{d\Theta}{d\ell} \right|_\su 
	= -\frac{3a^2}{\ra^3} \sin\theta \cos\theta
\end{equation*}
are components of the unit vector field normal to $\su$.
We can manipulate Eq.~(\ref{discont2}) in order to obtain
a more enlightening expression.
For this purpose, we make use of Eqs.~(\ref{ansatz})-(\ref{rel_coef}), 
as well as Taylor expansions of 
$d \chi_{\omega' l}(\rho) /d \rho |_{\rho = \rho_\su}$
and $d \psi_{\omega lm}(r) /d r |_{r = r_\su}$
around $\rho = \ra$ and $r = \ra$
[analogous to Eqs.~(\ref{chi_norm}) and~(\ref{psi_norm})].
Then, by multiplying Eq.~(\ref{discont2}) by $P_{l'}^m(\cos\theta)$ 
and integrating over $\theta$, we can cast the resulting equation
in the following form:
\begin{align} \label{discont3}
	& \beta_0^l N_{\omega' l m}^{-} + \frac{a^2}{M^2} 
	 \left[ \beta_1^l N_{\omega' l m}^{- (0)} 
	+ \beta_2^l N_{\omega' (l+2) m}^{- (0)} \right.
	\nonumber \\
	&\left. + \beta_3^l N_{\omega' (l-2)\, m}^{- (0)} 
	H(l - |m| - 2) \right] = 0, \qquad l \geq |m|,
\end{align}
where $\beta_j^l$, $j \in \{0,1,2,3\}$, are coefficients
which depend in principle on all mode and spacetime parameters
except on $a$ (and which we avoid writing
explicitly because of space restrictions). 
Then, in zeroth order in the 
rotation parameter, Eq.~(\ref{discont3}) reduces to
\begin{equation}
	\beta_0^l N_{\omega' l m}^{-(0)} = 0,
\end{equation}
which gives rise to a nontrivial solution for $\phi^-_{\omega'm}$
[see Eq.~(\ref{v_in_2})] if
\begin{equation} \label{beta0}
\beta_0^{l_0} = 0
\end{equation}
for some $l = l_0 \geq |m|$. In this case,
$N_{\omega' l_0 m}^{-(0)} \neq 0$ is fixed by the Klein-Gordon
normalization while $N_{\omega' l m}^{-(0)} = 0$ for $l\neq l_0$:
\begin{equation}
N_{\omega' l m}^{-(0)} = N_{\omega' l_0 m}^{-(0)}\; \delta_{l\, l_0}.
\label{N}
\end{equation}
Condition~(\ref{beta0}) can be written as
\begin{equation} \label{xi0}
	\xi^{(0)} =  \left( \frac{4 \mathsf{F}}{\ra}
	+ \frac{2M}{\mathsf{F} \ra^2} - \frac{4}{\ra} \right)^{-1}
	\left( \left. \frac{d \chi_{\omega' l_0}^{(0)}}{d\rho}\right|_{\ra}
	- \mathsf{F} \left. 
	\frac{d\psi_{\omega l_0 m}^{(0)}}{dr} \right|_\ra
		\right),
\end{equation}
which expresses the value of $\xi$ that the field must have
in order that unstable modes with quantum numbers $\omega'$ 
($=\omega/F$),
$l_0$ and $m$ do exist in the spacetime of a static
spherical shell with mass-to-radius ratio $M/\ra$.
In second order in $a/M$, Eq.~(\ref{discont3}) yields
\begin{align} \label{discont4}
	&\beta_0^l N_{\omega' l m}^{-(2)} 
	+ \beta_1^l N_{\omega' lm}^{-(0)} 
	 + \beta_2^l N_{\omega' (l + 2) \,m}^{-(0)} 
	\nonumber \\
	& + \beta_3^l N_{\omega' (l - 2) \,m}^{-(0)}
	H (l - |m| - 2) = 0, \quad l \geq |m|.
\end{align}
For $|m| \leq l \neq l_0$, Eq.~(\ref{discont4}) can 
be solved for $N_{\omega' lm}^{-(2)}$,
\begin{equation} \label{N-2}
	N_{\omega' l m}^{-(2)} = - N_{\omega' l_0 m}^{-(0)} \left(
	\frac{\beta_2^{l_0 - 2}}{\beta_0^{l_0 - 2}} 
	\delta_{l\, l_0 - 2}
	+ \frac{\beta_3^{l_0 + 2}}{\beta_0^{l_0 + 2}} 
	\delta_{l \, l_0 + 2}
	\right),
\end{equation}
which together with Eq.~(\ref{N}) determine $N_{\omega' l m}^-$ in Eq.~(\ref{N-})
(with $N_{\omega' l_0 m}^{-(0)}$ fixed by normalization).
Now, by using Eqs.~(\ref{N}) and~(\ref{N-2}) in Eq.~(\ref{discont3}),
we obtain
\begin{equation} \label{beta1}
\beta_0^{l_0}+ \frac{a^2}{M^2}	\beta_1^{l_0} = 0, 
\end{equation}
which can be explicitly written as
\begin{widetext}
\begin{align} \label{xi_eq}
	\xi 
	& \left( \frac{4\mathsf{F}}{\ra} + 
	\frac{2M}{\mathsf{F} \ra^2} 
	- \frac{4}{\ra} \right) \left[ 1 + \frac{a^2}{2\ra^2}
	\frac{4 \ra \mathsf{F}^4 - 4\ra \mathsf{F}^3 
	+ 4M \mathsf{F}^3 - M}{2 \ra \mathsf{F}^4 
	+ M \mathsf{F}^2 - 2\ra \mathsf{F}^3}
	+ \frac{a^2}{2 \ra^2} \frac{2l_0^2 + 2l_0 - 1 - 2m^2}
	{(2l_0 + 3)(2l_0 - 1)} \frac{12 \ra \mathsf{F} 
	- 12 \ra \mathsf{F}^2 - 9M \mathsf{F}^2}
	{2\ra \mathsf{F}^2 + M - 2 \ra \mathsf{F}} \right.
	\nonumber \\
	&\left. + \frac{a^2}{\ra} \left( \frac{M}{\ra} 
	- \frac{l_0^2 + l_0 - 3m^2}{(2l_0 + 3)(2l_0 - 1)}\right) 
	\left. \frac{d \chi^{(0)}_{\omega' l_0}}{d\rho} \right|_{\ra} 
	- \frac{a^2}{2\ra^2} \left(1 + \frac{2M}{\ra}\right)
	\frac{l_0^2 + l_0 - 3m^2}{(2l_0 + 3)(2l_0 - 1)}\right]
	= - \mathsf{F} \left. \frac{d\psi_{\omega l_0 m}}{dr} 
	\right|_\ra
	+ \left. \frac{d \chi_{\omega' l_0}}{d\rho}\right|_{\ra}
	\nonumber \\
	& - \frac{a^2 \mathsf{F}}{\ra^2} \left[ \alpha_1 \ra^2 
	- \frac{2l_0^2 + 2l_0 - 1 - 2m^2}{2(2l_0 + 3)(2l_0 - 1)}
	\left( 1 + \frac{2M}{\ra}\right)
	+\frac{1}{2\mathsf{F}^2}\right]	
	\left. \frac{d\psi^{(0)}_{\omega l_0 m}}{dr} \right|_\ra
	+ \frac{a^2 \mathsf{F}^3}{\ra} 
	\frac{2l_0^2 + 2l_0 - 1 - 2m^2}{(2l_0 + 3)(2l_0 - 1)} 
	\left. \frac{d^2\psi^{(0)}_{\omega l_0 m}}{dr^2} \right|_\ra
	\nonumber \\
	& + \frac{a^2}{\ra} \left( \frac{M}{\ra} 
	- \frac{l_0^2 + l_0 - 3m^2}{(2l_0 + 3)(2l_0 - 1)}\right) 
	\left. \frac{d^2 \chi^{(0)}_{\omega' l_0}}{d\rho^2}
	\right|_{\ra} - \frac{a^2}{2\ra^2}
	\frac{l_0^2 + l_0 - 3m^2}{(2l_0 - 1)(2l_0 + 3)}
	\left[ \frac{4 \mathsf{F}}{\ra} - \frac{6}{\ra} 
	+ \left( 1 + \frac{2M}{\ra} \right)
	\left. \frac{d\chi^{(0)}_{\omega' l_0}}{d\rho}
	\right|_{\ra} \right].
\end{align}
\end{widetext}
If we write $\omega = \omega_R + i \omega_I$, then, 
for each set $\{\omega_I, l_0, m, M/\ra, a/M\}$ of parameters,
Eq.~(\ref{xi_eq}) is a complex equation, the imaginary 
part of which can be solved for $\omega_R$, and the real 
part of which then returns a value for $\xi$. Then,
for each fixed $l_0 \geq |m|$, $\xi$
is the field coupling that marks the appearance of 
unstable terms (``partial modes'') in the sum in 
Eq.~(\ref{v_in_2}) with quantum numbers $\omega'$ and $m$,
and with $N^-_{\omega' l m}$ given by Eq.~(\ref{N-}) 
with Eqs.~(\ref{N}) and~(\ref{N-2}) [the corresponding
term in the exterior of $\su$ is determined from 
Eq.~(\ref{v_out}) together with Eqs.~(\ref{ansatz})
and~(\ref{rel_coef})].
The above procedure to calculate $\xi$ relies on the knowledge of
$\psi_{\omega lm}(r)$, which we compute numerically by
integrating Eq.~(\ref{psi2_eq}) subject to the boundary
condition (\ref{decay}) and normalization condition
$\psi_{\omega lm}(\ra) = 1$ [see discussion below Eq.~(\ref{psi_norm})]. 
In Fig.~\ref{fig:omega_var} this method is employed 
to obtain the values of $\xi$ and $M/\ra$ which 
trigger the instability for $a/M=0.2$, $l_0=m=0$ and 
$0<\omega_I\leq 0.4$. 
\begin{figure}[htb]
\includegraphics[width=8cm]{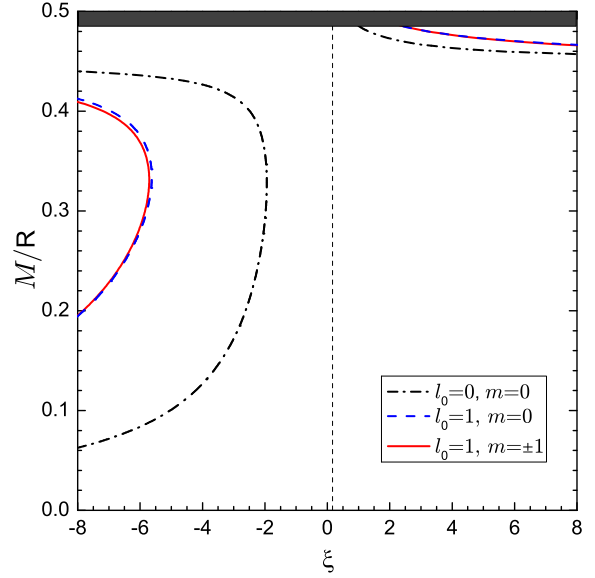}
\caption{Curves characterizing the onset of instability
for partial modes with $l_0=0$ and $l_0=1$ ($m=0$ and $m=\pm 1$).
The rotation is fixed to $a/M=0.2$. Curves for the same $l_0$ and
opposite values of $m$ are degenerate. 
The plot is restricted to the range of $M/\ra$ in which 
the criterion~(\ref{criterion}) is valid.
The vertical dashed line indicates the
conformal-coupling value $\xi = 1/6$. }
\label{fig:l_var}
\end{figure}
We note from Fig.~\ref{fig:omega_var} that the external boundaries
of the unstable regions (black curves) are numerically consistent with 
$\omega_I = \omega_R = 0$, which is compatible with the
general result \cite{Friedman1975} that instability sets 
in through zero-frequency modes~\cite{figuras}.
Therefore, let us now specialize to $\omega=0$,
in which case we can obtain analytically the second order
approximation for the implicit functions in Eq.~(\ref{xi_eq}).
Thus, we write up to second order
\begin{equation} \label{chi0}
	\chi_{\omega'_0 l}(\rho) = \chi_{\omega'_0 l}^{(0)}(\rho) 
	+ \frac{a^2}{M^2} \chi_{\omega'_0 l}^{(2)} (\rho),
\end{equation}
where $\omega'_0 \equiv - m \tilde{\Omega} / \mathsf{F}$ 
is the $\omega'$ frequency when $\omega=0$ and
\begin{equation} \label{psi0}
	\psi_{0 l m}(r) = \psi_{0 l m}^{(0)}(r) + 
	\frac{a^2}{M^2} \psi_{0 l m}^{(2)}(r).
\end{equation}
The static limit is straightforward:
\begin{equation} \label{chi_psi_0}
	\chi_{\omega'_0 l}^{(0)}(\rho) = \frac{\rho^l}{\ra^l}, 
	\qquad
	\psi_{0lm}^{(0)}(r) = \frac{Q_l(r/M - 1)}{Q_l(\ra /M - 1)},
\end{equation}
where $Q_l(x)$ is the Legendre function of the second kind.
The functions $\chi_{\omega'_0 l}^{(2)}(\rho)$ and
$\psi_{0 l m}^{(2)}(r)$ satisfy inhomogeneous differential
equations, for which the homogeneous part may be solved 
in terms of simple special functions. Therefore, 
standard methods (see, e.g. Ref.~\cite{Hassani1999}) can be 
used in order to derive the full expressions (\ref{chi0}) 
and (\ref{psi0}). In particular, we obtain
\begin{equation} \label{dchi0}
	\left. \frac{d\chi_{\omega'_0 l}}{d\rho}\right|_\ra = 
	\left. \frac{d\chi^{(0)}_{\omega'_0 l}}{d\rho}\right|_\ra 
	-\frac{4 a^2 M^2 m^2}{\ra^5 \mathsf{F}^2 (3 + 2l)}
\end{equation}
and
\begin{align} \label{dpsi0}
	\left. \frac{d\psi_{0 l m}}{dr} \right|_\ra 
	& = \left. \frac{d\psi^{(0)}_{0 l m}}{dr}\right|_\ra 
	+ \frac{a^2 C_l}{M^2 Q_l(\ra/M-1)} 
	\left[ \left. \frac{dP_l(r/M - 1)}{dr} \right|_\ra
	\right. \nonumber \\
	& \left. 
	- \frac{P_l(\ra /M - 1)}{Q_l(\ra/ M - 1)}
	\left.\frac{dQ_l(r/ M - 1)}{dr} \right|_\ra
	\right],
\end{align}
where
\begin{align} \label{Cl}
	C_l &\equiv \int_1^\infty{dx Q_l(x\ra/M -1) \left[
	\frac{M}{\ra} \left.\frac{d^2Q_l(t\ra /M - 1)}{dt^2} 
	\right|_{t=x} \right. }
	\nonumber \\
	& \left. + \frac{m^2}{x^2 \ra/M - 2 x}
	Q_l (x \ra / M - 1) \right].
\end{align}

\begin{figure}[htb]
\includegraphics[width=8cm]{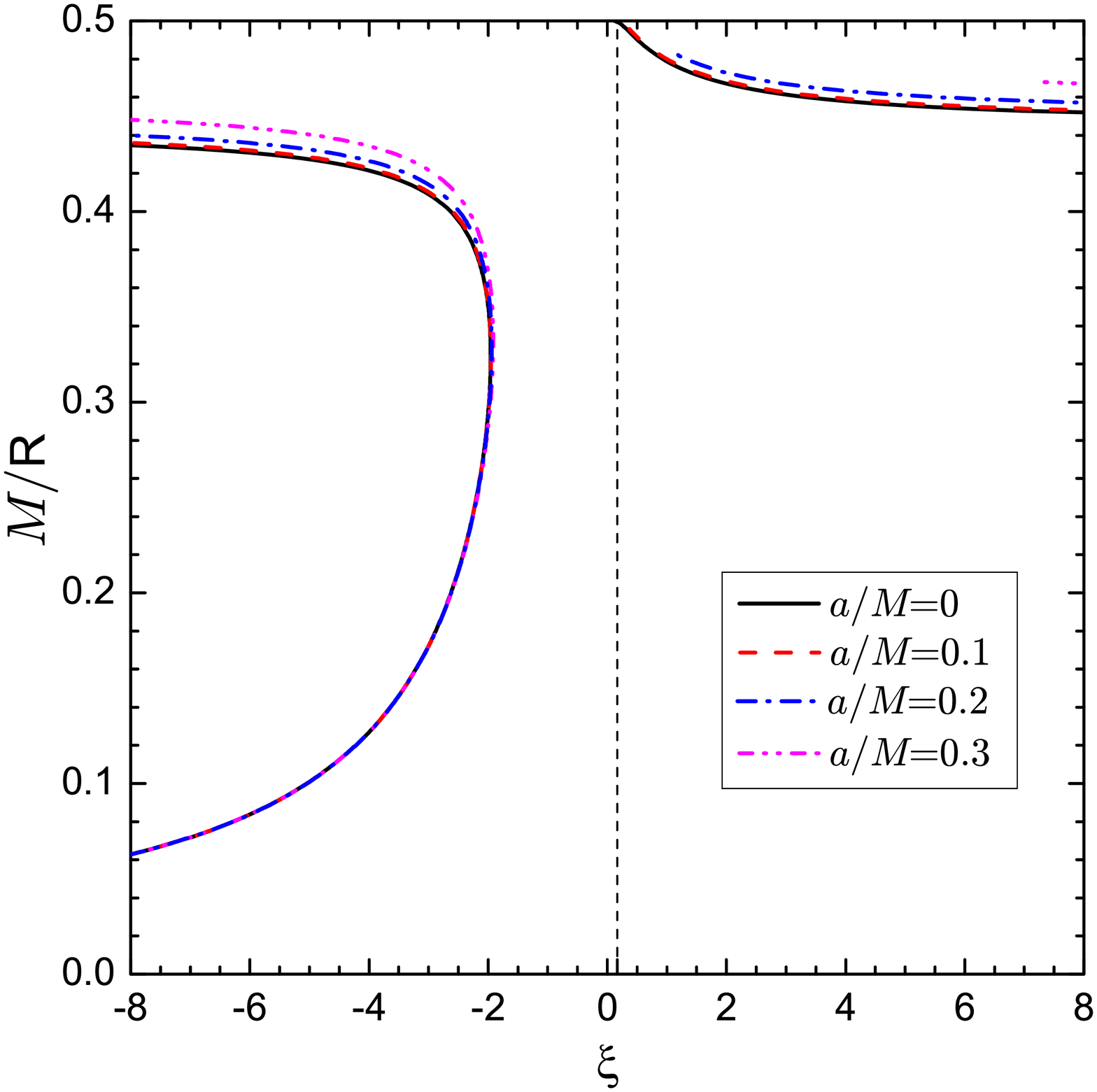}
\caption{Diagram showing the boundaries ($\omega = 0$, $l_0=0$)
of the regions in the parameter space $(\xi,M/\ra)$ in 
which the instability is triggered for $a/M= 0.1, 0.2,$ and 
$0.3$. The static $a=0$ case is plotted for comparison. 
The vertical dashed line indicates the conformal-coupling 
value $\xi = 1/6$. Configurations allowing for tachyonic 
modes are those to the left of the curves on the left-hand 
side and to the right of those on the right-hand side.
The curves are restricted to the range of $M/\ra$ in which 
the criterion~(\ref{criterion}) is valid.}
\label{fig:a_var}
\end{figure}
\begin{figure}[htb]
\includegraphics[width=8.8cm]{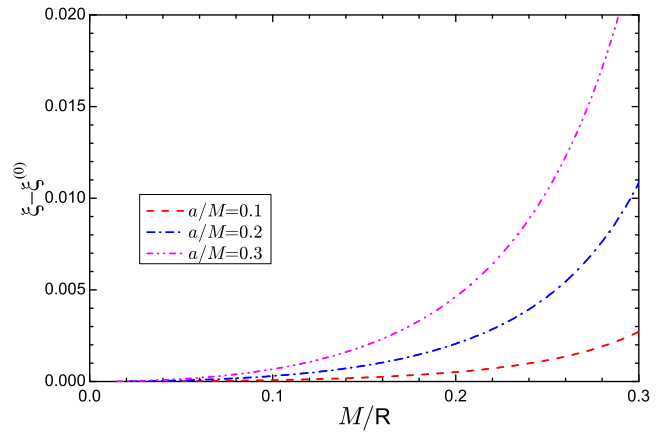}
\caption{Difference $\xi-\xi^{(0)}$ between the
values of $\xi$ describing the borders of the
unstable regions for a rotating shell and
for a static configuration with the same mass-to-radius
ratio, as a function of $M/\ra$.
Here, $l_0=m=0$ and $a/M = 0.1,0.2$, and $0.3$.
We see from the plot that for these values of $a/M$, 
$(\xi-\xi_0)/\xi_0 \lesssim (a/M)^4$, which suggests that 
for  $0 \leq M/R \lesssim 0.3$, a higher order analysis 
would be necessary to extract reliable conclusions.
}
\label{fig:xi2_l0}
\end{figure}

Therefore, by plugging Eqs.~(\ref{chi_psi_0}), 
(\ref{dchi0}), and (\ref{dpsi0}) into Eq.~(\ref{xi_eq}), 
we obtain an analytical expression [except for the simple
integral of Eq.~(\ref{Cl})] which can be directly solved
for $\xi$, giving the boundaries of the regions in the
parameter space where the instability sets in. 
Figure~\ref{fig:l_var} represents these limiting
curves for $a/M=0.2$ and different values of $l_0$ and $m$.
Clearly, the unstable regions for partial modes with $l_0=m=0$ 
encompass those for higher multipoles.

In Fig.~\ref{fig:a_var}, the boundaries ($\omega = 0$, $l_0=0$)
of the unstable regions are plotted for different values 
of $a/M$. Fig.~\ref{fig:xi2_l0} highlights a range of $M/\ra$ 
which is not clearly seen in Fig.~\ref{fig:a_var}. There,
we plot the difference $\xi-\xi^{(0)}$ as a function of $M/\ra$,
where $\xi^{(0)}$ and $\xi$ are given in Eqs.~(\ref{xi0}) 
and Eq.~(\ref{xi_eq}), respectively. From Figs.~(\ref{fig:a_var}) and
(\ref{fig:xi2_l0}), we conclude that rotation shifts 
these boundaries to the right, so that the unstable region 
for negative values of $\xi$ gets enlarged and the one for 
positive values of $\xi$ is diminished. The absolute effect, 
however, turns out to be relatively small (as expected, 
since this is a second order correction).

\begin{figure}[htb]
\includegraphics[width=8.8cm]{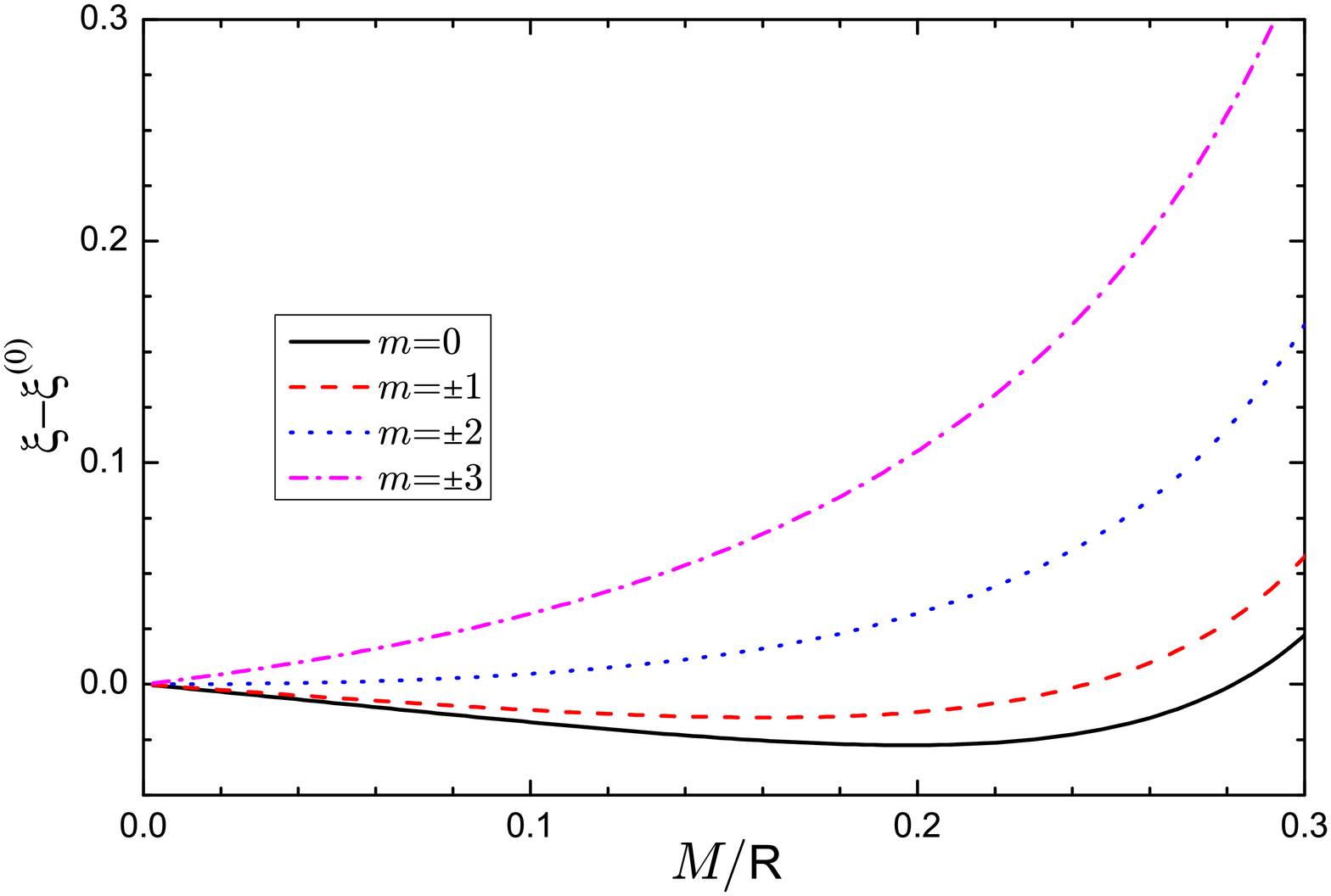}
\caption{Difference $\xi-\xi^{(0)}$ between the
values of $\xi$ describing the borders of the
unstable regions for a rotating shell and
for a static configuration with the same mass-to-radius
ratio, as a function of $M/\ra$.
Here, $a/M=0.3$, $l_0=3$ and $m=0,\pm 1, \pm 2$, 
and $\pm 3$.}
\label{fig:xi2_l3}
\end{figure}

Finally, Fig.~\ref{fig:xi2_l3} shows the value of $\xi-\xi^{(0)}$ as 
a function of $M/\ra$ for $a/M=0.3$, $l_0=3$, and 
$m=0,\pm 1,\pm 2, \pm 3$. Unlike the $l_0=0$ case, 
it turns out that for higher multipoles $\xi-\xi^{(0)}$ is
not everywhere positive. Here and in Fig.~\ref{fig:l_var} 
it is clear that a reversal in the direction of rotation 
(achieved either by $a \to -a$ or by $m \to -m$) has no 
effect in the parameter space of the instability, which 
is a direct consequence of the general result of 
Sec.~\ref{sec:1storder}. There is nonetheless a coupling
between the object rotation and the field angular momentum in higher
orders [manifested, e.g., by the term $a^2 m^2$ 
in Eq.~(\ref{psi2_eq})] which breaks the degeneracy in $m$
which is characteristic of the static limit.

Before concluding, it is worthwhile to make a brief comment 
on the relation between our results concerning the 
linear instability of nonminimally coupled fields 
in the spacetime of rotating bodies and the nonlinear 
effect known as spontaneous scalarization, which was 
established in the context of scalar-tensor theories 
in Ref.~\cite{Damour1993} (see also Refs.~\cite{Novak1998,Salgado1998,Sotani2012}). 
In Ref.~\cite{Pani2011}, it was argued that 
the boundaries of the regions in parameter space 
which characterize the type of linear instability 
considered here also delimit the regions where 
spontaneous scalarization can occur. Although the argument 
was made in a context of spherical symmetry, the same
reasoning seems to apply to our stationary spacetime. 
Indeed our conclusions are in agreement with
a recent result \cite{Doneva2013}, which numerically
showed that scalarized rapidly rotating neutron stars exist
for a larger range of (negative) couplings than in
the static case
(see Figs.~\ref{fig:a_var} and~\ref{fig:xi2_l0}).
\section{Conclusions}
\label{sec:conclusion}

Nonminimally coupled free scalar fields are unstable 
in the spacetime of compact objects for a wide range
of field couplings and compact object parameters. Such an
instability will be unavoidably triggered by vacuum fluctuations 
(see Sec.~\ref{sec:quantization}). This ``awakening" of the
quantum vacuum was previously treated in the context of 
spacetimes which were static in the asymptotic past and future
associated with the formation of a nonrotating compact  
object from initially diluted matter.
Here, we have investigated how the instability is 
influenced when the compact object acquires some rotation. 
In order to also allow  a quantum mechanical treatment of the 
instability, we have discussed the canonical quantization of 
the scalar field in a spacetime which is nearly flat in the 
asymptotic past and stationary and axisymmetric in the
future. As a prototype model 
for our compact spinning object, we have considered the spacetime 
of a spinning thin shell. As explained in Sec.~\ref{sec:1storder},
in order to obtain nontrivial results concerning the role of
rotation on the instability parameter space we had to go
beyond first order in the object angular momentum 
(see also Sec.~\ref{sec:2ndorder}). The simple 
thin shell model is justified, thus, since it allowed us
to push the analytical treatment further.
Our main result is expressed in Eq.~(\ref{xi_eq}) and
depicted in Figs.~\ref{fig:omega_var}-\ref{fig:xi2_l3}. 
In particular, we observe that the regions in parameter space 
which characterize the instability of a partial mode with a certain 
value of $l_0$ are invariant under $m \to - m$ but are 
nondegenerate in $m$ as can be seen in Figs.~\ref{fig:l_var} 
and~\ref{fig:xi2_l3} (in contrast to the static case).
Figure $3$ also shows that the instability first sets in by
partial modes with $l_0=0$.
Our analysis suggests that the overall effect of (slow) rotation 
is to enlarge the instability parameter space for negative
values of $\xi$ and to diminish the one for positive values 
of $\xi$ (see Figs.~\ref{fig:a_var} and~\ref{fig:xi2_l0}), 
in agreement with recent results in the context
of scalar-tensor theories \cite{Doneva2013}.

\acknowledgments
R. M. was supported by the S\~ao Paulo Research 
Foundation (FAPESP) under the Grant No. 2011/06429-3. 
G. M. and D. V. acknowledge partial support from FAPESP under
Grants No. 2007/55449-1 and 2013/12165-4, respectively.
G. M. also acknowledges Conselho Nacional de Desenvolvimento 
Cient\'\i fico e Tecnol\'ogico (CNPq) for partial support.


\begin{thebibliography}{}
\bibitem{figuras}
In our figures higher order contributions will be 
also present but a number of numerical tests leads 
us to believe that they will not change our physical 
conclusions.
\end{thebibliography}
\end{document}